\documentclass[twocolumn,showpacs,preprintnumbers,amsmath,amssymb]{revtex4}

\usepackage{graphicx}
\usepackage{dcolumn}
\usepackage{bm}

\begin{document}


\title{Current dependence of the 'red' boundary of superconducting single photon detectors in the modified hot spot model}

\author{D. Yu. Vodolazov$^{1,2}$}
\email{vodolazov@ipmras.ru} \affiliation{ $^1$ Institute for
Physics of Microstructures, Russian Academy of Sciences, 603950,
Nizhny Novgorod, GSP-105, Russia \\
$^2$ Lobachevsky State University of Nizhny Novgorod, 23 Gagarin
Avenue, 603950 Nizhny Novgorod, Russia}

\date{\today}


\begin{abstract}

We find the relation between the energy of the absorbed photon and
the threshold current at which the resistive state appears in the
current-carrying superconducting film with the probability about
unity. In our calculations we use the modified hot spot model,
which assumes different strength of suppression of the
superconducting order parameter in the finite area of the film
around the place where the photon is absorbed. To find the
threshold current we solve the Ginzburg-Landau equation for
superconducting order parameter, which automatically includes the
current continuity equation and it allows us to consider the back
effect of current redistribution near the hot spot on the
stability of the superconducting state. We find quantitative
agreement with the recent experiments, where we use the single
fitting parameter which describes what part of the energy of the
photon goes for the local destruction of the superconductivity in
the film.

\end{abstract}

\maketitle

\section{Introduction}

In 2001 it was demonstrated that the current-carrying
superconducting film biased near its critical current can detect
single photon in the visible and near infrared range of
electromagnetic radiation (act of detection was seen in the
experiment as an appearance of the voltage pulse in the film)
\cite{Gol'tsman}. The one of the main characteristics of such a
superconducting single photon detector (SSPD) is the position of
'red' boundary, or the largest wavelength of the photon
($\lambda_{max}$) which could be detected with probability close
to unity. In reality, the part of photons can miss the active
element of SSPD - superconducting film, which automatically makes
its system detection efficiency smaller than unity. This problem
has a technical character and may be solved, for example, by
putting a superconducting film in the form of meander to the
resonator \cite{Marsili}. To characterize the intrinsic ability of
superconducting film to detect the photons it was offered the new
characteristic - intrinsic detection efficiency (IDE)
\cite{Hofherr}, which has a meaning of probability of the photon
detection after its absorption {\it by the superconducting film}
(from this definition it follows that IDE $\simeq 1$ when $\lambda
\leq \lambda_{max}$).

In many experiments it was observed that $\lambda_{max}$ depends
on the transport current (see for example
\cite{Semenov_hot_spot,Maingault,Engel_APL,Lusche}). Moreover at
fixed $\lambda$ IDE does not drop suddenly when current becomes
smaller some critical (threshold) value but it decreases gradually
until it reaches unmeasurably small values. At the moment there
are several theoretical models
\cite{Semenov_hot_spot,Maingault,Zotova,Bulaevskii} which relates
the threshold current $I_{thr}$ with $\lambda_{max}$. These models
are based on the assumption that the absorbed photon creates the
hot spot
\cite{Semenov_hot_spot,Maingault,Zotova,Engel_JAP,Semenov_first,Eftekharian}
or the hot belt \cite{Bulaevskii} in the superconducting film (for
their comparison see Ref. \cite{Lusche}). In the present work we
use a bit different model for the hot spot (see section 2) than
the one used in Ref. \cite{Zotova}. To study the situation when
IDE$\simeq 1$ we place the hot spot in the center of the film and
identify the critical current of such a system as a threshold
current. This assumption is based on our recent study
\cite{Zotova2} where we find that with current increase IDE
gradually increases from $\sim 0.05$ (when only the photons
absorbed near the edge of the film provide the instability of
superconducting state) up to the unity (when the instability
arises also from the photons absorbed in the center of the film).
We make our calculations for the films with different width and
the hot spots with different size, which correspond to the photons
with different energies. We compare our results with recent
experiments \cite{Engel_APL,Lusche,Renema_PRB} and find good
quantitative agreement with the only assumption that about $10 \%$
of the photon energy goes for the local suppression of the
superconductivity.

\section{Model}

We consider a two-dimensional superconducting film with finite
width and the hot spot is modelled by the local area (in the form
of the circle - see insets in Fig. 1(a)) where the quasiparticle
distribution function $f(\epsilon)$ is supposed to be far from the
equilibrium. We assume that this nonequilibrium is created by the
photon and it affects the stability of the superconducting state
with transport current. To find the value of the critical current
of the film with such a hot spot we numerically solve the
Ginzburg-Landau equation for the superconducting order parameter
$\Delta=|\Delta|e^{i\varphi}$
\begin{equation}
\xi_{GL}^2\left(\frac{\partial^2\Delta}{\partial
x^2}+\frac{\partial^2\Delta}{\partial
y^2}\right)+\left(1-\frac{T_{bath}}{T_c}+\Phi_1-\frac{|\Delta|^2}{\Delta_{GL}^2}\right)\Delta=0
\end{equation}
with the additional term \cite{Ivlev,Larkin,Kramer}
\begin{equation}
\Phi_1=\int_{|\Delta|}^{\infty}\frac{2(f^0-f)}{\sqrt{\epsilon^2-|\Delta|^2}}de,
\end{equation}
which describes the effect of nonequilibrium distribution function
$f(\epsilon)\neq f^0(\epsilon)=1/(exp(\epsilon/k_BT_{bath})+1)$.
In Eq. (1) $\xi_{GL}^2=\pi\hbar D/8k_BT_c$ and
$\Delta_{GL}^2=8\pi^2(k_BT_c)^2/7\zeta(3)\simeq 9.36(k_BT_c)^2$
are the zero temperature Ginzburg-Landau coherence length and the
order parameter correspondingly. Note, that imaginary part of Eq.
(1) leads to the current continuity equation $div j_s=0$ ($j_s$ is
a superconducting current density) which allows us to find the
proper distribution of the current density in the film with the
hot spot.

It is convenient to write Eq. (1) in dimensionless units (length
is scaled in units of
$\xi(T_{bath})=\xi_{GL}/(1-T_{bath}/T_c)^{1/2}$ and $|\Delta|$ in
units of $\Delta_{eq}=\Delta_{GL}(1-T_{bath}/T_c)^{1/2}$)
\begin{equation}
\frac{\partial^2\tilde{\Delta}}{\partial
\tilde{x}^2}+\frac{\partial^2\tilde{\Delta}}{\partial
\tilde{y}^2}+\left(\alpha-|\tilde{\Delta}|^2\right)\tilde{\Delta}=0,
\end{equation}
where $\alpha=(1-T_{bath}/T_c+\Phi_1)/(1-T_{bath}/T_c)$. In Eq.
(3) the impact of the absorbed photon on the superconducting
properties of the film is described by the single parameter
$\alpha$, which in reality should vary in time and space due to
$f(\epsilon,\vec{r},t)$ (in equilibrium $\alpha=1$). In our model
we use the static approach for the hot spot (with
$\alpha(t)=const$ inside the hot spot). We can do it because of
different time scales existing in this problem. Indeed, during
initial very short time the hot quasiparticles (appearing after
photon absorption - see for example \cite{Kozorezov}) undergo the
downconvesrion cascade to the energy level just above $\Delta$ and
the further relaxation process develops much longer. The low
energy nonequilibrium quasiparticles diffuse in space (which is
relatively slow process due to small group velocity of
quasiparticles seating at the energies close to the energy gap),
but simultaneously they suppress $\Delta$ locally below its
equilibrium value. It results in the appearance of quasiparticles
with the energy less than $\Delta_{eq}$ which already cannot
diffuse and they become trapped by the hot spot (see more detailed
discussion of this phenomena in Ref.\cite{Kozorezov}). These
quasiparticles can relax to the equilibrium only via
electron-phonon scattering with characteristic inelastic
electron-phonon relaxation time $\tau_{e-ph}$. Therefore roughly
this time governs the final stage of evolution of the hot spot.
But the time change of $\Delta$ is much faster process and at low
temperatures it is proportional to $\hbar/\Delta_{eq}$ which is
much shorter than $\tau_{e-ph}$. Therefore on the time scale of
change of $\Delta$ one may consider the quasiparticle distribution
function as a static object (at the final stage of its evolution).

In our previous paper \cite{Zotova} in the numerical simulations
we used the model with the local temperature of the quasiparticles
and solved the heat conductance equation for space and temporal
evolution of $T_{loc}$. This model oversimplifies the real
situation because it does not take into account the aforementioned
suppressed diffusion of the quasipartices with $\epsilon \lesssim
\Delta_{eq}$ and, besides, it assumes implicitly that at any
moment in time the quasiparticles are in the local equilibrium
which is rough approximation in the real SSPD where inelastic
electron-electron relaxation time is comparable with inelastic
electron-phonon relaxation time. But nevertheless it gives {\it
qualitatively} the same results as using here the 'static' hot
spot model. It is not surprising, because in the case when
$f(\epsilon)$ is described by the Fermi-Dirac function with the
local temperature $T_{loc}$ the our control parameter
$\alpha(\vec{r},t)=(1-T_{loc}(\vec{r},t)/T_c)/(1-T_{bath}/T_c)$.
Physically it corresponds to the step like distribution of
$T_{loc}$ in space which provides qualitatively similar
suppression of $\Delta$ as it does the Gaussian-like distribution
of $T_{loc}$ following from the heat conductance equation (see for
example Eq. (13) in Ref. \cite{Zotova}).

Note, that different $f(\epsilon)$ may provide the same value of
$\alpha$, which is consequence of dependence of superconducting
order parameter $\Delta$ on integral of $f(\epsilon)$ (with some
weight function) over the energy. Finding $f(\epsilon)$ needs the
solution of the kinetic equation which is very difficult problem
\cite{Kozorezov} and it is beyond the scope of this paper. But to
have an insight on the possible values of $\alpha$ let as assume
for simplicity that $f(\epsilon)$ is described by the Fermi-Dirac
function with the local temperature $T_{loc}$. Due to diffusion of
hot quasiparticles the region where $T_{loc}>T_{bath}$ grows but
the most effectively the order parameter is suppressed in the area
where $T_{loc}>T_c$ and $\alpha<0$. At some moment, after the
absorption of the photon, the region where $T_{loc}>T_c$ stops to
grow and one can use this moment for determination of the
effective size of the hot spot.
\begin{figure}[hbtp]
\includegraphics[width=0.52\textwidth]{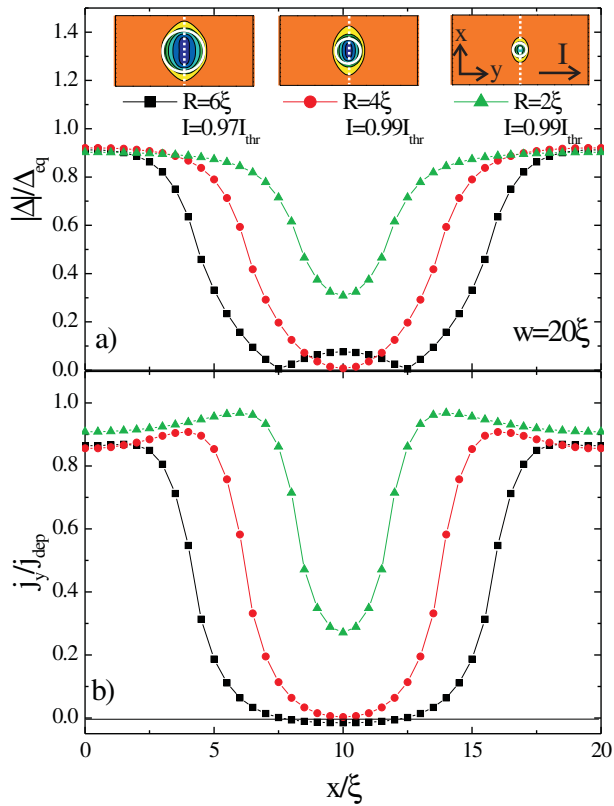}
\caption{Distribution of $|\Delta|$ (a) and superconducting
current density (b) along the white lines marked in the insets
(where we plot contour plot of $|\Delta|$ in the film). The region
where we put $\alpha=0$ is bordered by the white circle in the
insets in figure (a) and at all radiuses the transport current is
close to the threshold value.}
\end{figure}

Based on above consideration we choose $\alpha=0$ inside the spot
in the form of circle with radius $R$ (we also checked what occurs
at different values of $\alpha$). The radius $R$ and the energy of
the photon $ch/\lambda$ are roughly related as
\begin{equation}
\eta \frac{ch}{\lambda}\simeq d\pi R^2\frac{H_{cm}^2}{8\pi}
\end{equation}
where $H_{cm}=\Phi_0/2\sqrt{2}\pi\xi\lambda_L$ is the
thermodynamic magnetic field, $\Phi_0$ is the magnetic flux
quantum, $\lambda_L$ is the London penetration depth, $d$ is the
thickness of the film and $H_{cm}^2/8\pi$ is the superconducting
condensation energy per unit of volume. Coefficient $0<\eta<1$
takes into account that not whole energy of the photon goes for
suppression of $\Delta$ \cite{Semenov_hot_spot}. For example the
large part of the energy of the photon is delivered to the energy
of quasiparticles $E_q$. One can easily find it in the local
temperature approximation (when $T_{loc}(\vec{r})=const$ inside
the circle with radius R)
\begin{eqnarray}
E_q=d\pi R^24N_0\int_{0}^{\infty}\epsilon f(\epsilon)d\epsilon=
\nonumber
\\
d\pi R^2N_0\frac{\pi^2k_B^2T_{loc}^2}{3}\simeq  d\pi
R^2N_0\Delta_0^2\frac{1.06T_{loc}^2}{T_c^2}
\end{eqnarray}
where $N_0$ is a one-spin density of states of quasiparticles at
the Fermi level, $\Delta_0=1.76 k_BT_c$ is the superconducting
order parameter at zero temperature and we neglect the energy of
quasiparticles in equilibrium (which is small in comparison with
Eq. (5) when $\Delta_{eq}>k_BT_{bath}$ and $T_{loc}\gg T_{bath}$).
One can see that when $T_{loc}=T_c$, $E_q$ is twice larger than
the condensation energy (because at low temperatures
$H_{cm}^2/8\pi \simeq N_0\Delta_0^2/2$) and when $T_{loc}=2T_c$,
$E_q$ is larger more than in eight times.

In numerical calculations we consider the film of finite width $w$
and length $L=4w$ with the hot spot (region where $\alpha<1$)
placed in the center of the film - see insets in Fig.1a. We also
add to the right hand side of Eq. (3) the term with time
derivative $\partial {\tilde \Delta}/\partial t$ (in a way how it
was done in Ref. \cite{Zotova}) which allows us to find not only
the value of the critical (threshold) current but also to find how
the superconducting state becomes destroyed. As an edge effect
during the nonstationary stage the normal current $j_n$ (and
electrostatic potential) appears in the film and we find them by
solving equation $div (j_n+j_s)=0$.

\section{Results}

In Fig. 1(a,b) we show distribution of $|\Delta|$ and current
density across the hot spot and sidewalks. Note that despite of
$\alpha=0$ the order parameter is finite in the hot spot region
(see Fig. 1(a)). It occurs due to proximity effect from the
surrounding area and the same effect leads to partial suppression
of $|\Delta|$ outside the region where we keep $\alpha$. Due to
the current crowding effect the maximum of current density is
located near the 'edge' of the hot spot (compare Fig. 1(a) and
1(b)). The same result follows from the analytical calculations in
the London model (see Ref. \cite{Zotova}). This result is opposite
to the findings of Ref. \cite{Engel_JAP} (see Fig. 5 there) where
authors find the maximal current density at the edge of the film.
We checked for different $w$ and $R$ that the current density is
maximal at the edge of the film only when the size of the hot spot
is comparable with the width of the film (one can see it on the
example of the hot spot with diameter $D=12\xi\sim w=20\xi$ in
Fig. 1(b)).

When the current exceeds some critical value the vortex-antivortex
pair appears {\it inside} the hot spot. It occurs due to large
value of the supervelocity ($v_s\simeq j_s/|\Delta|^2$) inside the
hot spot \cite{Zotova}, which actually affects the
superconductivity. For relatively large hot spots (with $R\gtrsim
3 \xi$) this pair is unbound at larger current, at which vortex
and antivortex move freely in the opposite directions. One can see
the signs of the presence of the vortex-antivortex pair from
distribution $|\Delta|(x)$ and $j_y(x)$ in Fig. 1(a,b) for the
spot with $R=6\xi$. Dependence $|\Delta|(x)$ has two local minima
near the center of the film which correspond to the centers of the
vortex and antivortex. From Fig. 1(b) one can see that $j_y$ is
negative near the center of the film which is the consequence of
the currents flowing around the vortex/antivortex centers and
which have opposite direction to the direction of the transport
current.
\begin{figure}[hbtp]
\includegraphics[width=0.52\textwidth]{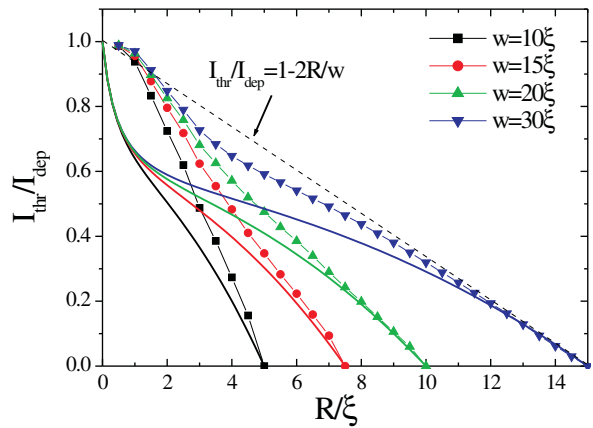}
\caption{Threshold current as a function of the radius of the spot
($\alpha=0$) for films with different widths. Solid curves
correspond to Eq. (12) from Ref. \cite{Zotova} with $\gamma=0$.
Dotted line corresponds to the dependence $I_{thr}(R)$ for the
film with width $w=30 \xi$ in the model with spatially uniform
current distribution in the sidewalks near the hot spot.}
\end{figure}

We relate the current, at which the free motion of vortices and
antivortices start, with the threshold current at which IDE of
SSPD approaches unity (in Ref. \cite{Zotova2} we find that when
the hot spot located {\it not} in the center of the film the free
vortex motion starts at {\it lower} currents). In Fig. 2 we
present dependence of $I_{thr}/I_{dep}$ on the radius of the spot
when $\alpha=0$. In the same figure we present dependence
$I_{thr}(R)$ (solid curves) following from the London model (see
Eq. (12) in Ref. \cite{Zotova}). The quantitative difference
between these results is not surprising, because in the analytical
model of Ref. \cite{Zotova} the gradual variation of $|\Delta|$ at
the edge of the hot spot was neglected and semi-quantitative
criteria for unbinding of vortex-antivortex pair was used.

In Fig. 2 we also plot $I_{thr}(R)$ (dashed line) for the film
with $w=30 \xi$ which follows from the model with spatially
uniform current distribution in the sidewalks \cite{Maingault}.
One can see that the current crowding effect leads to smaller
value of $I_{thr}$ when $\xi \ll R \ll w$.
\begin{figure}[hbtp]
\includegraphics[width=0.46\textwidth]{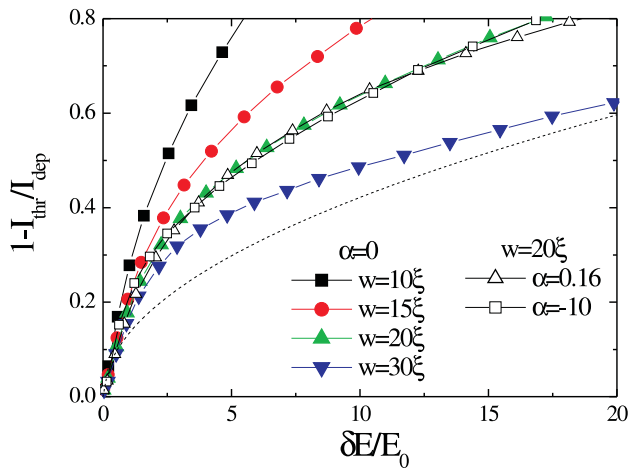}
\caption{Dependence of the threshold current on the energy
difference between the states of the superconducting film with and
without the hot spot (we did not take into account the difference
in the energy of quasiparticles).}
\end{figure}

In Fig. 3 we show the same dependencies but instead of radius of
the spot, as one of the coordinate, we use the difference between
the free energies of the states of the film with and without the
hot spot
\begin{equation}
\delta E=F_{spot}-F_{no spot}-\frac{\hbar}{2e}I\delta\varphi.
\end{equation}
where
\begin{equation}
F=-\frac{H_{cm}^2d}{8\pi}\int \left( \frac{|\Delta|}{\Delta_{eq}}
\right)^4 dS,
\end{equation}
and $\delta \varphi$ is the extra phase difference due to presence
of the hot spot. At the derivation of Eq. (7) from the
Ginzburg-Landau free energy functional we take into account that
$|\Delta|$ is the solution of the Ginzburg-Landau equation and we
do not consider difference in the energy of quasiparticles because
it strongly depends on explicit dependence $f(\epsilon)$. In Fig.
3 (where the energy is normalized in units of
$E_0=H_{cm}^2d\xi^2/2=\Phi_0^2d/16\pi^2\lambda_L^2$) for the film
with $w=20\xi$ we also show the results for the hot spots with
different values of $\alpha$. Surprisingly, the difference is not
large in the used coordinates (notice, that for larger $\alpha$
the same value $I_{thr}$ is reached for larger radius of the hot
spot, but the energies of these states are turned out to be close
to each other).

\section{Comparison with the experiment}

To relate the theoretical results present in Figs. 2,3 with the
recent experiments \cite{Engel_APL,Lusche,Renema_PRB} we suppose
that only fraction of the photon's energy $\eta hc/\lambda$ goes
for destruction of superconductivity inside the hot spot. In Figs.
4,5 we compare our results with the experimental results present
in Refs. \cite{Renema_PRB,Engel_APL,Lusche}. In Fig. 4 we use
$\xi=7 nm$ and the square resistance $R_{sq}=400 Ohm$ for all
theoretical curves. Depairing current for TaN meander from Ref.
\cite{Engel_APL} is calculated with the Kupriyanov-Lukichev
correction \cite{Kupriyanov}. In Ref. \cite{Engel_JAP} this
correction was omitted \cite{Engel_PC} and it leaded to slightly
smaller value of $I_{dep}$ (compare Fig. 4 with the inset in Fig.
9 of Ref. \cite{Engel_JAP}). In Ref. \cite{Renema_PRB} authors did
not present parameters of the NbN bridge ($R_s$ or $\lambda_L$ and
$T_c$) and we take these parameters from Ref. \cite{Kamlapure} for
NbN film with the same thickness as in Ref. \cite{Renema_PRB}.

In Figs. 4,5 the only fitting parameter is the coefficient $\eta$.
It has almost the same value about 0.1 for different TaN meanders.
For NbN bridge the best fitting is obtained for $\eta=0.17$
((note, that other hot spot models give $\eta\simeq 0.1-0.4$ after
their comparison with the experiments
\cite{Semenov_hot_spot,Engel_APL,Engel_JAP}). We should mention
that in Ref. \cite{Renema_PRB} the IDE was fixed at the level
$\sim 0.25$ (this number can be estimated using the experimental
results for detection probability $p_n$ in Fig. 3 of Ref.
\cite{Renema_PRB}). From Fig. 3 of Ref. \cite{Renema_PRB} one can
see that $p_n$ saturates (and IDE$\to 1$) at larger currents. It
means that the experimental points in Fig. 5 should be shifted
upwards and in this case the theoretical results would fit the
experimental ones practically at the same value of $\eta$ as for
TaN meanders.
\begin{figure}[hbtp]
\includegraphics[width=0.46\textwidth]{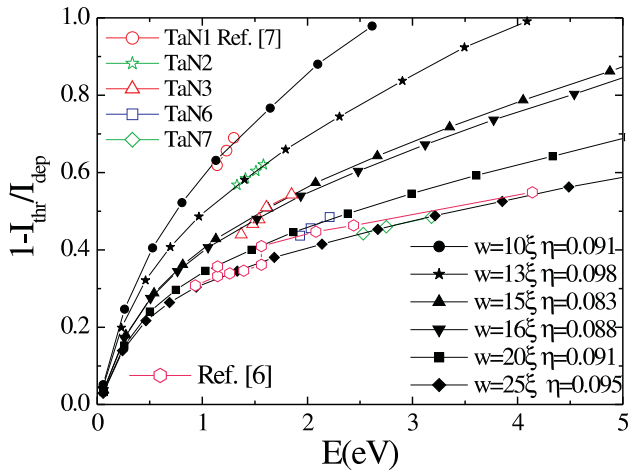}
\caption{Dependence of the threshold current, when IDE$\to 1$, on
the energy of the absorbed photon. The experimental results were
found for TaN based SSPD in Refs. \cite{Engel_APL,Lusche}.}
\end{figure}

\begin{figure}[hbtp]
\includegraphics[width=0.48\textwidth]{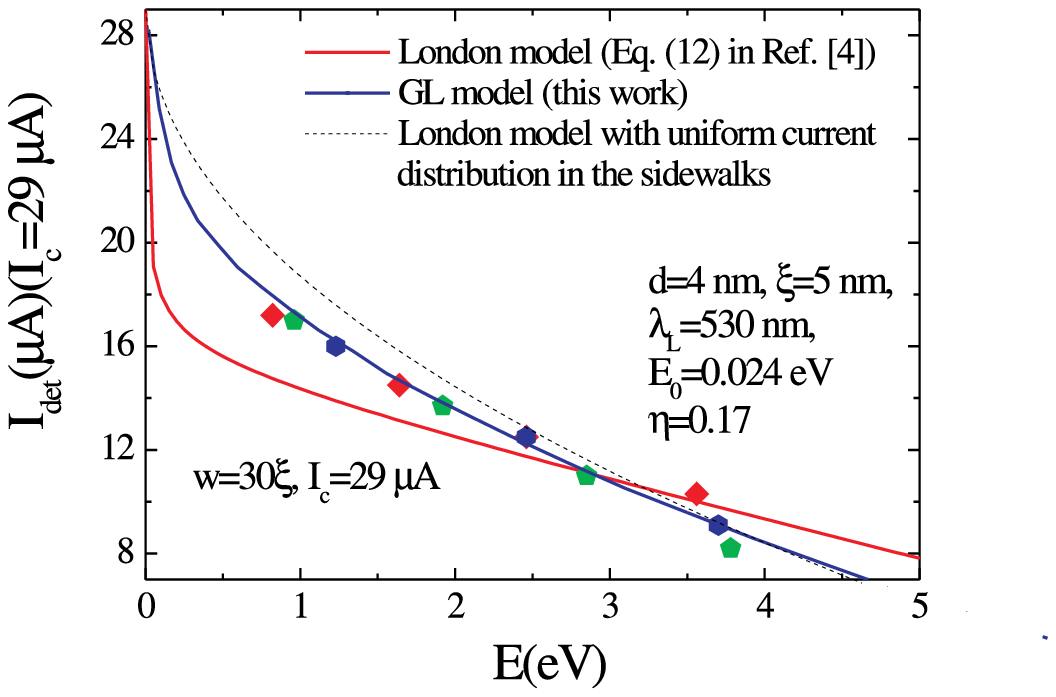}
\caption{Symbols - experimental results from Ref.
\cite{Renema_PRB} found when IDE$\sim 0.25$. Solid curves -
theoretical results calculated in GL and London models when
IDE$\simeq 1$. Dashed line corresponds to the model with uniform
current distribution in the sidewalks near the normal spot.}
\end{figure}

There is also another work \cite{Renema_PRL}, where the photon
detection by the NbN bridge (in the same geometry as in
\cite{Renema_PRB}) was studied. In this work authors fix the
detection probability at $p_n=0.01$ (while in the previous work
they used $p_n=0.1$ \cite{Renema_PRB}) and find linear dependence
of $I_{thr}$ on the energy of the absorbed photons, which provides
such value of $p_n$. Assuming the presence of correlation between
the studied bridges in Refs. \cite{Renema_PRB,Renema_PRL} one may
suggest that $p_n=0.01$ corresponds to IDE$\sim 0.025$. In this
case we cannot compare directly this experiment with our results
because they are obtained when IDE$\sim 1$.

Note, that if one extrapolates the experimental points in Figs.
4,5 (and in Fig. 2 of Ref. \cite{Renema_PRL}) by linear functions
then they cross the vertical axis at $I_{thr} \sim 0.7-0.8
I_{dep}$($I_c$). This circumstance leaded to the conclusion in
Refs. \cite{Engel_JAP,Renema_PRL} that the penetration of the
vortices via edge energy barrier is important for photon detection
process, because in Ref. \cite{Bulaevskii} it was found that this
barrier goes to zero at current smaller when $I_{dep}$. We should
stress here that the energy barrier in Ref. \cite{Bulaevskii} is
calculated in the London model and this approach quantitatively
fails when the vortex sits on the distance smaller than $\sim
2\xi$ from the edge of the film. Calculations in the framework of
the Ginzburg-Landau model \cite{Vodolazov_barrier} demonstrate
that the energy barrier for the vortex penetration vanishes {\it
exactly} at $I=I_{dep}$ and only at currents $I<0.6 I_{dep}$ the
London and the Ginzburg-Landau models give the same results (if
one takes into account the energy of the vortex core
\cite{Vodolazov_barrier}). Note, that if in the film there are
some geometric or structural inhomogeneities then the energy
barrier for the vortex entry vanishes at the critical current of
the film $I_c<I_{dep}$ \cite{Vodolazov_barrier}. In this case
IDE$\simeq 1$ could be reached only for the photons, those energy
satisfies the inequality $I_{thr}(E)\leq I_c$.

\section{Conclusion}

In this work we exploit the modified hot spot model to find the
relation between the energy of the absorbed photon and the
threshold current at which the resistive state appears {\it with
probability about unity} in the current-carrying superconducting
film. The hot spot is modelled as a finite region in the center of
the film where the quasiparticle distribution function is far from
the equilibrium and it leads to the partial suppression of the
superconducting order parameter. We find that the resistive state
starts from the nucleation and unbinding the vortex-antivortex
pair {\it inside} the hot spot. When the energy of the photon goes
to zero, the suppression of the order parameter in the hot spot
becomes small and the threshold current rapidly approaches the
depairing current. Comparison of found results with recent
experiments shows good agreement if one assumes that only about
$10\%$ of the energy of the photon goes for the local destruction
of the superconductivity.

The main differences of our model with the previous hot spot
models
\cite{Semenov_hot_spot,Maingault,Engel_JAP,Semenov_first,Eftekharian}
are following: i) we solve the current continuity equation $div
j_s=0$ in the film with the hot spot (which gives us correct
distribution of the current density in the superconductor); ii) we
take into account the back effect of the current redistribution
around the hot spot on the superconducting order parameter.
Because we did not solve the kinetic equation and we did not find
nonequilibrium distribution function we do not know the actual
value of $\alpha$ and we cannot relate quantitatively the energy
of the photon with the size of the hot spot and how strong the
superconductivity is suppressed inside it. But we find that the
relation between the threshold current and the energy needed for
suppression of the superconductivity inside and around the hot
spot (which does not include the energy of hot quasiparticles and
phonons) weakly depends on the these parameters. This result
encourage us that the found results are rather general and have
direct relation to the dependence of the 'red' boundary of SSPD on
the transport current.

\begin{acknowledgments}

We thank to Alexander Semenov, Gregory Gol'tsman, Alexey Semenov,
Jelmer Renema, Michiel de Dood and Martin van Exter for valuable
discussion of the found results. The work was partially supported
by the Russian Foundation for Basic Research (project 12-02-00509)
and by the Ministry of education and science of the Russian
Federation (the agreement of August 27, 2013,  N 02.Â.49.21.0003
between The Ministry of education and science of the Russian
Federation and Lobachevsky State University of Nizhni Novgorod).

\end{acknowledgments}

\end{document}